\documentclass[aps,prb,superscriptaddress,twocolumn,showpacs]{revtex4-1}
\usepackage{graphicx}
\usepackage{amsmath}
\usepackage{epsfig}
\usepackage{multirow}
\usepackage{array}
\usepackage{inputenc}
\usepackage{color}
\usepackage{tabularx}
\usepackage{textcomp}
\usepackage{ulem}
\usepackage{hhline,multirow,float}
\usepackage[colorlinks=true,linkcolor=black, citecolor=blue, urlcolor=blue,
    unicode=true]{hyperref}
\begin{document}


\title{Decoupled molecular and inorganic framework dynamics in CH$_3$NH$_3$PbCl$_3$}


\author{M. Songvilay}
\affiliation{School of Physics and Astronomy, University of Edinburgh, Edinburgh EH9 3FD, UK}
\author{Zitian Wang}
\affiliation{School of Physics and Astronomy, University of Edinburgh, Edinburgh EH9 3FD, UK}
\author{V. Garcia Sakai}
\affiliation{ISIS Facility, Rutherford Appleton Laboratory, Chilton, Didcot OX11 0QX, UK}
\author{T. Guidi}
\affiliation{ISIS Facility, Rutherford Appleton Laboratory, Chilton, Didcot OX11 0QX, UK}
\author{M. Bari}
\affiliation{Department of Chemistry and 4D LABS, Simon Fraser University, Burnaby, British Columbia, V5A1S6, Canada}
\author{Z.-G. Ye}
\affiliation{Department of Chemistry and 4D LABS, Simon Fraser University, Burnaby, British Columbia, V5A1S6, Canada}
\author{Guangyong Xu}
\affiliation{NIST Center for Neutron Research, National Institute of Standards and Technology, 100 Bureau Drive, Gaithersburg, Maryland, 20899, USA}
\author{K. L. Brown}
\affiliation{School of Physics and Astronomy, University of Edinburgh, Edinburgh EH9 3FD, UK}
\author{P. M. Gehring}
\affiliation{NIST Center for Neutron Research, National Institute of Standards and Technology, 100 Bureau Drive, Gaithersburg, Maryland, 20899, USA}
\author{C. Stock}
\affiliation{School of Physics and Astronomy, University of Edinburgh, Edinburgh EH9 3FD, UK}

\date{\today}


\begin{abstract}

The organic-inorganic lead halide perovskites are composed of organic molecules imbedded in an inorganic framework.  The compounds with general formula CH$_{3}$NH$_{3}$PbX$_{3}$ (MAPbX$_{3}$) display large photovoltaic efficiencies for halogens $X$=Cl, Br, and I in a wide variety of sample geometries and preparation methods.  The organic cation and inorganic framework are bound by hydrogen bonds that tether the molecules to the halide anions, and this has been suggested to be important to the optoelectronic properties.  We have studied the effects of this bonding using time-of-flight neutron spectroscopy to measure the molecular dynamics in CH$_3$NH$_3$PbCl$_3$ (MAPbCl$_3$).  Low-energy/high-resolution neutron backscattering reveals thermally-activated molecular dynamics with a characteristic temperature of $\sim$ 95\,K.  At this same temperature, higher-energy neutron spectroscopy indicates the presence of an anomalous broadening in energy (reduced lifetime) associated with the molecular vibrations.  By contrast, neutron powder diffraction shows that a spatially long-range structural phase transitions occurs at 178\,K (cubic $\rightarrow$ tetragonal) and 173\,K (tetragonal $\rightarrow$ orthorhombic).  The large difference between these two temperature scales suggests that the molecular and inorganic lattice dynamics in MAPbCl$_3$ are actually decoupled.  With the assumption that underlying physical mechanisms do not change with differing halogens in the organic-inorganic perovskites, we speculate that the energy scale most relevant to the photovoltaic properties of the lead-halogen perovskites is set by the lead-halide bond, not by the hydrogen bond.

\end{abstract}


\maketitle


\section{Introduction}

The lead halide perovskites have generated considerable interest within the condensed matter physics community because they exhibit outstanding photovoltaic and optoelectronic properties that have already been exploited in the design of new solar cell devices.~\cite{Egger2016,Egger2018,Brenner2016,Saparov2016,Jeon2015,Shi2018,Zhu2019} These materials can be grouped into two categories: the hybrid organic-inorganic perovskites and the all-inorganic perovskites. The first group comprises materials based on an inorganic framework of PbX$_6$ octahedra (X = I, Br, Cl) and an organic molecular cation (commonly methylammonium -MA- or formamidinium -FA- ).~\cite{Kieslich2014}  Although the molecular cation does not directly contribute to the optoelectronic properties, its asymmetric shape and its ability to rotate and vibrate within the interstices of the inorganic framework affects the crystal structure and, thus, the electronic band gap.~\cite{Bakulin2015,Saparov2016,Motta2015,Warwick2019}  Furthermore, quasielastic neutron scattering studies of MAPb(Br,I)$_3$ \cite{Brown2017,Swainson2015,Leguy2015} demonstrate that the molecular cation participates in an order-disorder transition that is also associated with a dielectric anomaly and may therefore be connected with changes in the photoluminescence spectra as a function of temperature.~\cite{Wright2016}  However, while MAPbI$_3$ shows the highest photovoltaic power conversion efficiency to date (over 25\%)~\cite{NREL,Chen2019} among the lead halide perovskites,~\cite{Weller2015,Saparov2016,Ren2016} recent experimental studies have shown that all-inorganic perovskite materials actually display comparable efficiencies (15\% for CsPbI$_3$).~\cite{Wang2019}  Consequently, the role and importance of the organic cation to the photovoltaic properties remains unclear.~\cite{Lee2017,Kulbak2015}

In order to understand the interplay between the organic cation and the inorganic framework, as well as the effect of the molecule on the thermal conductivity and optoelectronic properties, numerous studies have been devoted to characterizing the molecular dynamics in the lead halides using optical spectroscopy techniques (Raman, infrared), x-ray and neutron scattering methods, and NMR.~\cite{Maalej1997,Quarti2014,Leguy2016bis,Mosconi2014,Bakulin2015,Baikie2015,Brivio2015} Some of these works also address the coupling between the relaxational dynamics arising from the organic cation and the phonons associated with the inorganic framework in both powder and single crystal samples.~\cite{Brown2017,Swainson2015,Songvilay2018,Leguy2015}  All these studies observe an onset of fast and overdamped (i.\ e.\ short-lived) relaxational molecular dynamics on heating that coincides with a distortion in the inorganic unit cell and are indicative of a coupling between the inorganic lattice and molecular dynamics.

\begin{figure}
\includegraphics[scale=0.6]{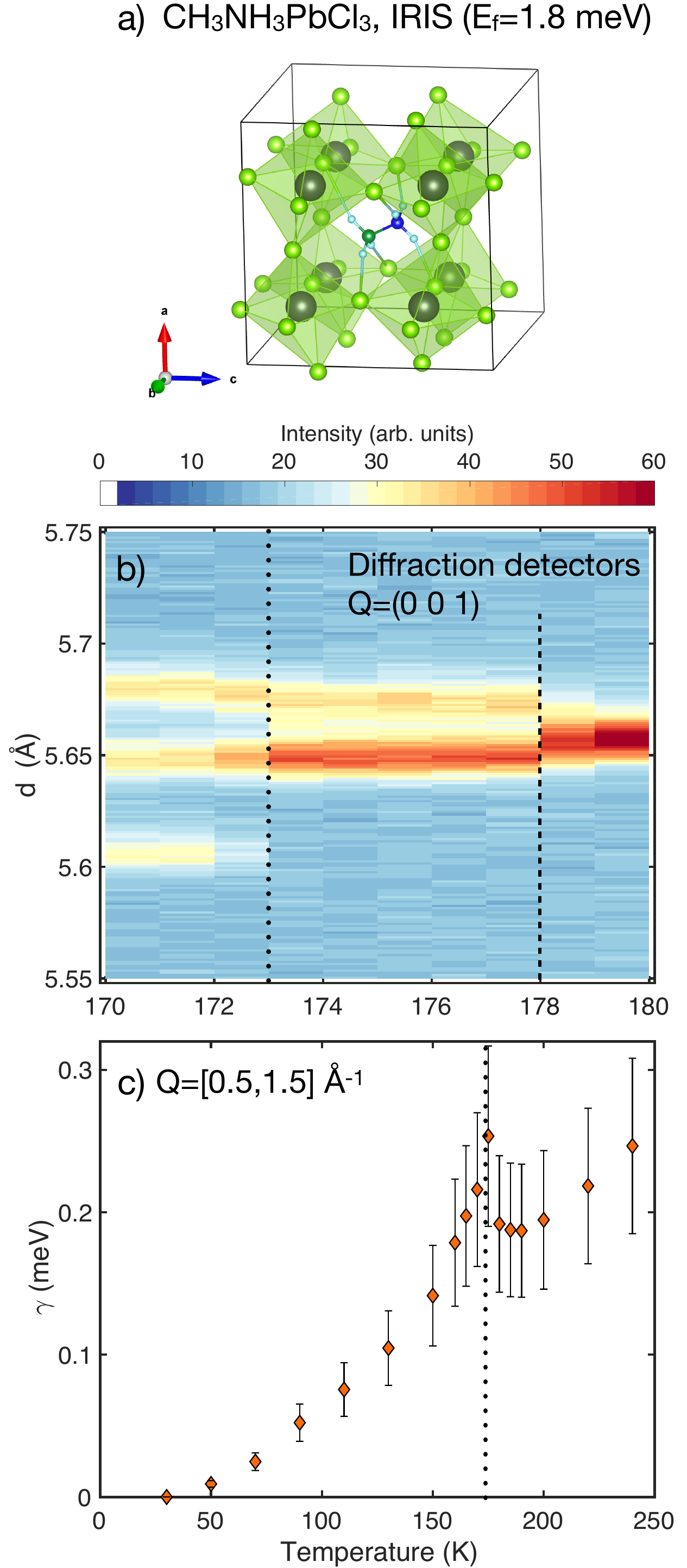}
\caption{ \label{fig:structure} $(a)$ Crystallographic structure of MAPbCl$_3$ in the orthorhombic phase. The Pb cations are represented by grey spheres and the Cl anions by green spheres. In this ordered phase, the H, N, and C atoms forming methylammonium molecules are represented with light blue, dark blue, and dark green spheres, respectively. $(b)$ Temperature dependence of the intensity of the (0~0~1) Bragg reflection, measured with the diffraction detectors of IRIS. The dashed lines denote transitions to the tetragonal (178\,K) and orthorhombic phases (173\,K). $(c)$ Temperature dependence of the quasi-elastic linewidth. The dashed line indicates the order-disorder transition.}
\end{figure}

MAPbCl$_3$ (Fig.~\ref{fig:structure} $(a)$) undergoes two structural transitions on cooling: a cubic-to-tetragonal transition at 178\,K and a transition to an orthorhombic phase at 173\,K.  This sequence of transitions is reflected in the temperature dependence of the ${\bf{Q}}$=(0~0~1) Bragg peak intensity shown in Figure \ref{fig:structure} $(b)$, measured on the IRIS neutron backscattering spectrometer, and is similar to that observed in the bromine and iodine analogues. These phases were first characterized by Poglitsch and Weber~\cite{Potglisch1987} who identified the respective space groups as cubic $Pm\overline{3}m$ ($a$ = 5.675 \AA), tetragonal $P4/mmm$ ($a$ = 5.656 \AA, $c$ = 5.630 \AA), and orthorhombic $P222_1$ ($a$ = 5.673 \AA, $b$ = 5.628 \AA, $c$ = 11.182 \AA).  Chi \textit{et al.} \cite{Chi2005} later suggested that the orthorhombic space group was actually $Pnma$ ($a$ = 11.1747 \AA, $b$ = 11.3552 \AA, $c$ = 11.2820 \AA), based on powder diffraction measurements. In this low temperature $Pnma$ phase it was shown that the methylammonium cations order in an antiparallel arrangement, along with a strong distortion of the PbCl$_6$ octahedra, mediated by the hydrogen bonds between the methylammonium and the chlorine anions. That the octahedral distortion occurs upon ordering of the molecules is evidence that the molecular cation is the most rigid unit in this compound.~\cite{Chi2005}  A single crystal diffraction study of the intermediate tetragonal phase was also carried out by Kawamura and Mashiyama,~\cite{Kawamura1999, Swainson2005} who observed both superlattice and incommensurate reflections, which were also reported for MAPbBr$_3$ \cite{Guo2017} and MAPbI$_3$.~\cite{Jacques2019}

In this paper, we investigate the dynamics of MA molecular cations in MAPbCl$_3$ using neutron inelastic scattering and provide a comparison with previous results on other halide materials. As the halogen radius affects the inorganic framework size, the radius of the space available for the MA molecule to rotate is also expected to change depending on the halogen anion, thus affecting the dynamics and therefore potentially the optoelectronic properties. We will show that there is a decoupling between molecular and framework dynamics and discuss the role of the molecule in the electronic and low-energy framework dynamics.

\section{Experimental details}

Powder samples were prepared out of solution following the procedure outlined in Ref.~\onlinecite{Chi2005}.   Methylamine hydrochloride and lead acetate were dissolved in hydrochloric acid.  An excess of approximately 8-10 molar ratio of methylamine hydrochloride was required to obtain phase pure samples.  The resulting powder was washed multiple times with diethyl ether to remove residual acid.  For all neutron measurements the powder was wrapped in niobium foil as a precaution to prevent reaction of the aluminum sample holder with any residual acid on the powder.

High resolution measurements of the quasielastic scattering were performed on the IRIS spectrometer (ISIS, UK).  IRIS is an indirect geometry, time-of-flight spectrometer that uses PG002 crystal analyzers to produce a fixed final neutron energy $E_f$=1.84\,meV.  The energy transfer is given by $E = E_i - E_f$, with a default dynamic range of $\pm$ 0.5\,meV, and the spectrometer energy resolution, given by the full-width at half maximum (FWHM) of a Gaussian fit at low temperatures, was $\delta E$=23\,$\mu$eV.   An empty can measurement for background subtraction was carried out at room temperature.  The IRIS spectrometer provides a diffraction bank of detectors at $2\theta$ $\sim$ 170$^{\circ}$ that allowed us to measure the structural transitions and unit cell parameters simultaneously while also measuring the excitations.

The high-energy molecular dynamics was studied using the MARI direct geometry spectrometer (ISIS, UK).  The incident neutron energy was fixed at $E_i=100$, 50, and 25\,meV using a Gd Fermi chopper spinning at a frequency of 600\,Hz, 400\,Hz, and 400\,Hz respectively.  This configuration gave energy resolutions at the elastic position ($E=0$) of 2.25, 1.15, and 0.48\,meV respectively.  The incident energies were small enough that hydrogen recoil scattering is not relevant for the discussion here.~\cite{Stock10:81}  A thick disc chopper was used and spun at 50\,Hz to remove high energy neutrons before the Fermi chopper.  Neutrons with very high energy ($\sim$ eV) were filtered with a $t0$ chopper spun at 50\,Hz.  For all measurements on MARI and IRIS, a closed-cycle helium refrigerator was used to control the temperature of the sample.

\section{Neutron spectroscopy}

Our goal in performing neutron spectroscopic measurements on the powders of MAPbCl$_3$ was to track the temperature dependence of the molecular dynamics so that we could correlate the response to distortions of the inorganic framework and crystal structure and compare our findings to previous studies of the bromine variant.  To this end we conducted experiments on two spectrometers, each designed to cover a substantially different energy range and therefore probing different timescales.  The IRIS backscattering spectrometer, which provides extremely sharp energy resolution, was used to characterize the quasielastic scattering, which is observed in systems that display diffusive behavior.  Quasielastic scattering is low-energy inelastic scattering that peaks at zero energy transfer ($E=0$) and has a non-zero intrinsic energy width.  In the case of MAPbCl$_3$, quasielastic scattering is sensitive to the re-orientational motions or jumps of the MA molecule within the interstices of the inorganic framework.  We emphasize here that such re-orientational dynamics are not harmonic modes which would instead appear as peaks at non-zero energy transfer.  The time-of-flight chopper instrument MARI, which provides access to dynamics at much higher energy transfers, was then used to study the higher-energy harmonic molecular modes and their temperature dependence.

\subsection{Quasielastic neutron scattering}

We first discuss the quasielastic scattering, and in this context it is important to note that the neutron incoherent cross section of hydrogen is roughly one order of magnitude larger than the cross sections (coherent or incoherent) of the other atoms in MAPbCl$_3$.  Specifically, the neutron incoherent cross section for hydrogen is 80.3\,barns whereas the next largest coherent/incoherent scatterer is chlorine, which has a cross section of only 11.5\,barns/5.3\,barns.~\cite{Sears1992}  For this reason, the quasielastic scattering we observed is effectively a measure of the low-energy molecular dynamics associated with single-particle motions of the hydrogen atoms.

\begin{figure}
 \includegraphics[scale=0.44]{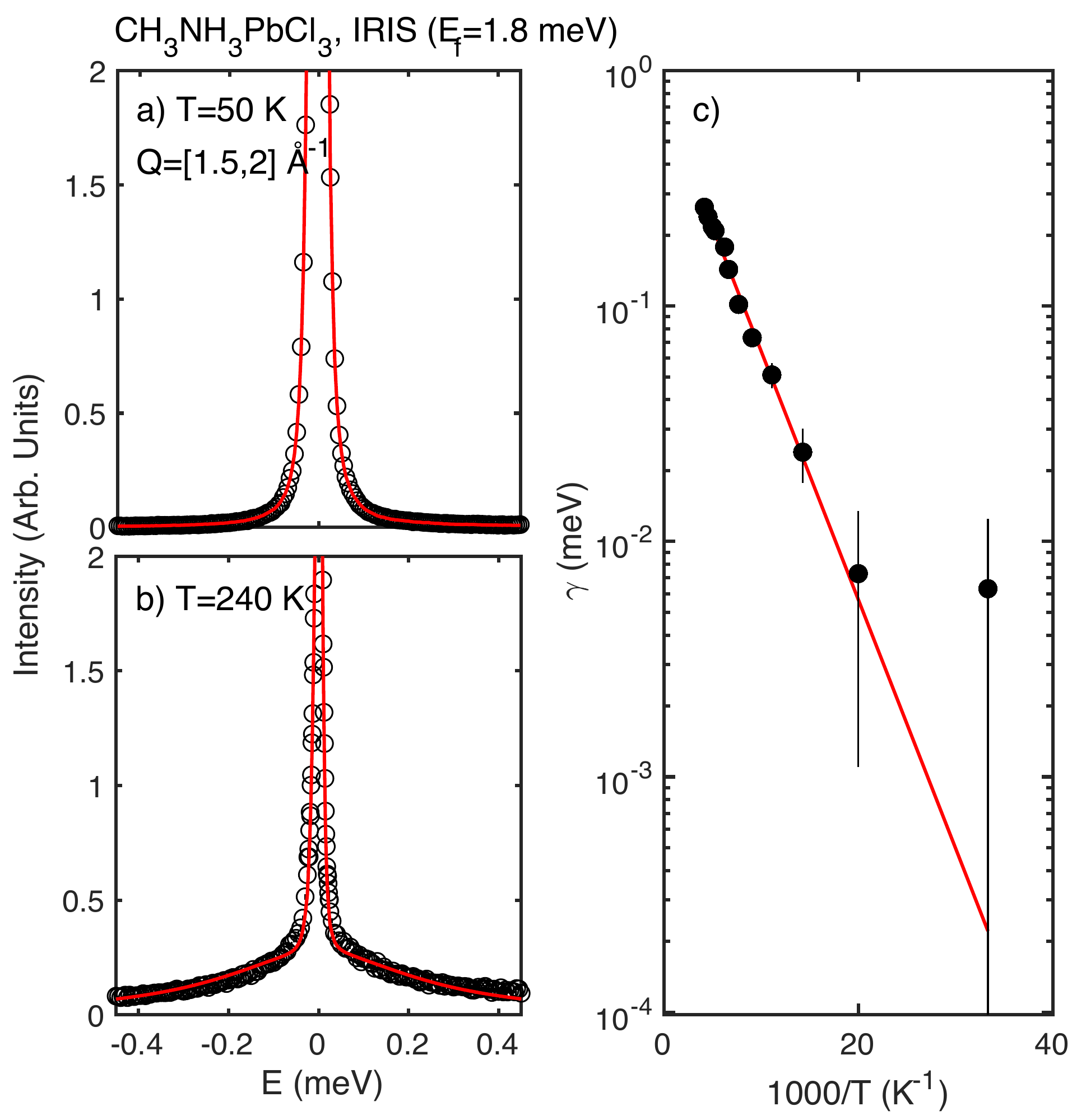}
 \caption{ \label{fig:figure_E_depend} $(a,b)$ Quasielastic scans measured on the IRIS backscattering spectrometer at 50\,K and 240\,K.  The scan at 50\,K is dominated by a strong resolution-limited peak at $E=0$, which indicates that the molecular dynamics are static on the timescale set by the instrumental resolution.  The scan at 240\,K exhibits an energy profile with two distinct components, both centered at $E=0$: one with a resolution-limited energy width (like that at 50\,K but weaker), and another that is much broader in energy, which reflects the molecular dynamics.  The solid curves are fits to the model discussed in the text.  $(c)$ Plot of Arrhenius fits to the quasistatic energy linewidths as a function of temperature.}
\end{figure}

In order to motivate an appropriate choice of scattering cross section $\tilde{I}(Q,E)$ with which to model the neutron quasielastic scattering from MAPbCl$_3$, we begin by examining two spectra measured at low (50\,K) and high (240\,K) temperatures, shown in Fig.~\ref{fig:figure_E_depend} $(a)$ and $(b)$, respectively, that were obtained by integrating the scattered intensity over $Q$ from 1.5\,\AA\ to 2.0\,\AA.  The 50\,K spectrum is well described by a Gaussian function of energy with a width equal to the instrumental elastic resolution, which means that no dynamics are visible at 50\,K on the timescale sampled by IRIS.  This spectrum changes on heating to 240\,K, as shown in Fig.~\ref{fig:figure_E_depend} $(b)$, where it evolves into a two-component energy profile.  The first component exhibits a resolution-limited energy width like that at 50\,K, but it is noticeably weaker; by contrast, the second component is much broader and extends out to at least $\pm$ 0.5\,meV.  The presence of two components suggests that the molecular dynamics of MAPbCl$_3$ can be characterized by two timescales, one of which is static compared to the dynamic range accessible by IRIS, and the other of which is dynamic and inversely proportional to the energy width of the broad component.  Thus, to parameterize the quasielastic scattering in terms of the momentum transfer $Q$, energy transfer $E$, and temperature $T$ our data were fit to the following model scattering cross section that contains both a static and a dynamic component:

\begin{equation}
\begin{split}
\tilde{I}(Q,E)=I_{el} (Q)+I_{dyn} (Q) \\
=I_{el} (Q) \delta(E) + {I_{dyn} (Q) \over {1+(E/\gamma(T))^{2}}}.
\label{lineshape}
\end{split}
\end{equation}

\noindent Here, $I_{el}$ and $I_{dyn}$ are the static and dynamic components, respectively, $\gamma$ is the energy linewidth, which is inversely proportional to the lifetime of the molecular fluctuations $\tau \sim 1/\gamma$, and $\delta(E)$ is a Dirac delta function in energy.  This cross section was convolved with the spectrometer resolution function given by the 50\,K spectrum as mentioned above, and representative fits to the data are shown as solid red curves in Figs.~\ref{fig:figure_E_depend} $(a,b)$.  For a given temperature, we find that the energy linewidth $\gamma$ is independent of momentum transfer $Q$, which indicates that no measurable diffusion of molecules occurs in the sample.  We also note that the single Lorentzian used to describe the dynamic component in Fig.~\ref{fig:figure_E_depend} $(b)$ shows small, systematic deviations from the data at small energy transfers.  This may indicate the presence of multiple timescales and a need for a more complicated model cross section.  The values for $\gamma$ that we obtain from our fits should therefore be interpreted as an average or a range of timescales rather than that for a single damped harmonic oscillator.  Regardless, our use of a single Lorentzian function and a single dynamical timescale describes the data quite well over a broad range of temperatures and wave vector $Q$ and thus captures the most important physical features of the low-energy molecular dynamics.

\begin{figure}
 \includegraphics[scale=0.37]{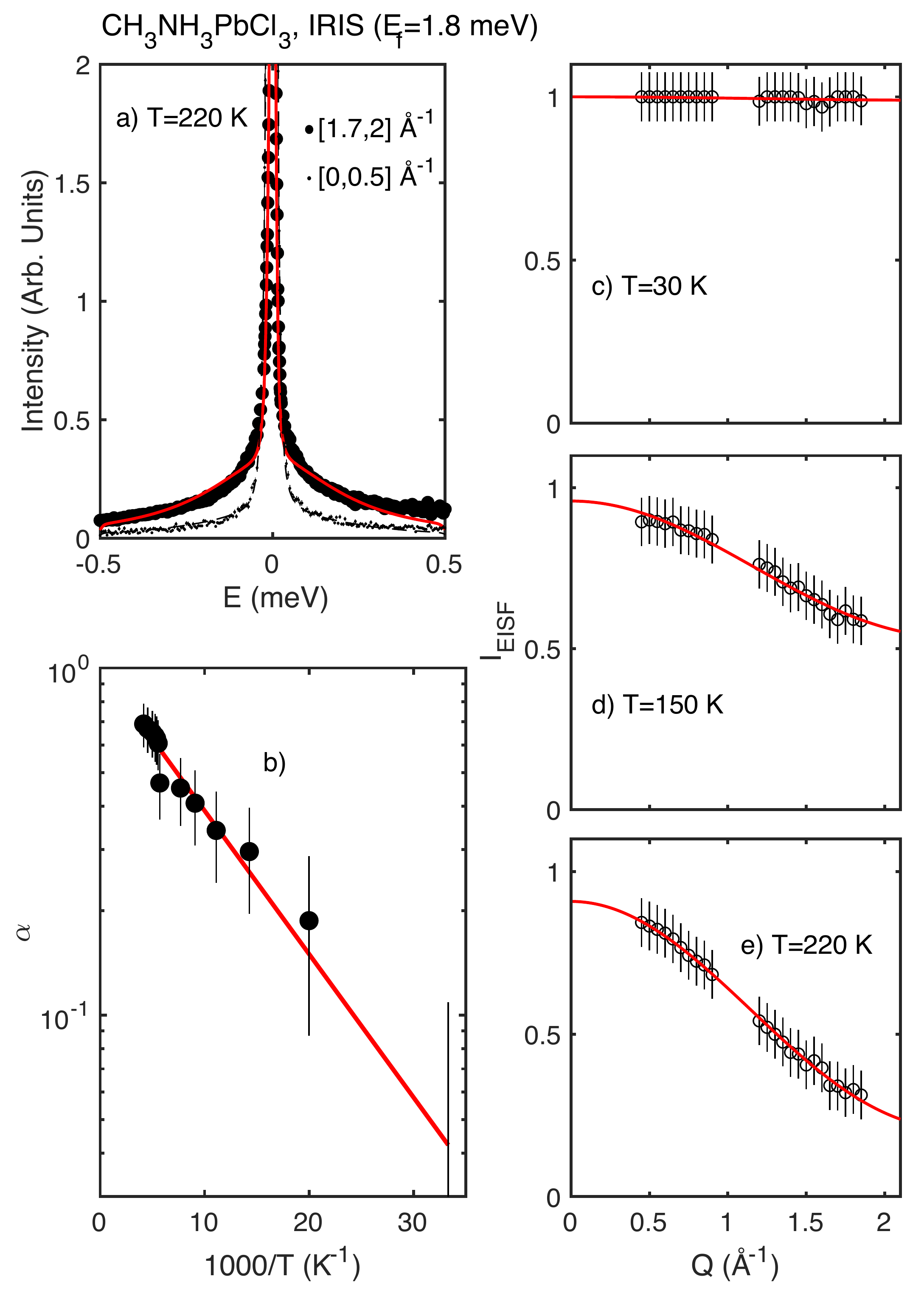}
 \caption{ \label{fig:evol_temp} $(a)$ Energy dependence of the quasielastic scattering at 220\,K for two different $Q$-ranges.  $(b)$ Arrhenius plot of the dynamic molecular fraction determined by the EISF discussed in the text. $(c-e)$ Plots of the EISF at 30\,K, 150\,K, and 220\,K, respectively.  The solid curve is a fit to a two-component model consisting of static/bound and dynamic molecules.}
\end{figure}

The temperature dependence of the quasielastic linewidth $\gamma$ is shown in Fig.~\ref{fig:structure} $(c)$, and a substantial increase is observed in the vicinity of the two structural transitions.  We note that due to the small temperature difference between the two structural transitions, we were not able to definitively observe these separate transitions in the molecular dynamics.  This behavior demonstrates the presence of some coupling between the inorganic framework and molecular dynamics, which then gives rise to critical scattering near the transitions.  The strength of this coupling can be quantified by fitting the temperature dependence of $\gamma$ to an Arrhenius law as shown in Fig.~\ref{fig:figure_E_depend} $(c)$, where $\gamma$ is plotted on a log scale as a function of $1000/T$.  The data within $\pm 10$\,K of the structural transitions have been removed because they are dominated by the critical scattering.  The red solid line in Fig.~\ref{fig:figure_E_depend} $(c)$ is a fit of the remaining data to the expression $\gamma(T) = \gamma_0\exp(-E_a/k_BT)$, where $\gamma_0 = 8.5 \pm 0.8$\,K and $E_a = 243 \pm 15$\,K.  These parameters are smaller than those obtained for the bromine variant (see Ref.~\onlinecite{Swainson2015}), where $\gamma_0(Br) = 27 \pm 2$\,K and $E_a(Br) = 323 \pm 20$\,K, and indicates that the molecular dynamics in MAPbCl$_3$ are softer than that in MAPbBr$_3$.  Recent neutron studies of the iodine variant report an even larger activation energy $E_a(I) \sim 600$\,K~\cite{Chen2015,Leguy2015}, although a more complex rotational jump model was used to analyze the data.  Nevertheless, the results on MAPbX$_3$ establish a clear trend in which the molecular dynamics hardens as the halide changes from Cl to Br and to I.  It is tempting to attribute this trend to a halogen-dependent variation in the strength of the N-H-X hydrogen bonding between the inorganic framework and the molecular cation and this is corroborated by calculations for hydrogen bonding with Halogens~\cite{Brammer01:1} and consistent with the idea that the framework structure is strongly correlated with the molecule.~\cite{Lee16:28} But first-principles calculations of the hydrogen-bonding strength in MAPbX$_3$ yield a value of 0.09\,eV/cation that is identical for X = Cl, Br, and I~\cite{Svane2017} and indeed the necessity of the molecule for the exceptional semiconducting properties has been questioned.~\cite{Zhu17:29}  If this is correct, then there must be some other bonds that set the energy scale of the molecular dynamics in the organic lead halide perovskites.

Our discussion so far has dealt primarily with the energy dependence of the quasielastic scattering.  We now move on to examine the momentum dependence, which provides information about the real-space symmetry of the molecular dynamics in which the hydrogen atoms participate through the elastic incoherent structure factor (EISF), which is defined as

\begin{equation}
\begin{split}
I_{EISF}(Q)= {I_{el(Q)} \over {I_{el}(Q)+I_{dyn}(Q)}}.
\end{split}
\end{equation}

\noindent Different model cross sections have been used successfully to describe a wide variety of molecular dynamics, including isotropic rotational dynamics and rotational jump dynamics, each of which exhibits a different dependence on momentum transfer.  But the data must span a sufficiently large $Q$-range in order to distinguish between these models and determine the symmetry of the dynamics in MAPbCl$_3$ uniquely.  This is because the EISF for these models is nearly identical for $Q$ less than $\sim$ 2\,\AA$^{-1}$, a point that is illustrated in Figure~3 of Ref.~\onlinecite{Brown2017}.  Neutron scattering kinematics limits the maximum $Q$ accessible on the IRIS backscattering spectrometer to less than $\sim$ 2\,\AA$^{-1}$.  For this reason we have chosen the simplest model, which describes the molecular dynamics in terms of two components, one static, corresponding to bound molecules, and the other dynamic.  This later component corresponds to unbound molecules which contribute to the inelastic signal fitted to a lorentzian lineshape decribed above.  Thus the total neutron scattering cross section has the form

\begin{equation}
\begin{split}
I_{total}(Q)=(1-\alpha) I_{bound} +  \alpha \tilde{I} \\
 \equiv (1-\alpha) I_{bound} +  \alpha [I_{el} (Q)+I_{dyn} (Q)],
\end{split}
\end{equation}

\noindent where $(1-\alpha)$ is the fraction of bound molecules that do not contribute to the dynamics.  A similar model was applied in Ref.~\onlinecite{Line1994} to characterize the dynamics of water molecules in minerals, and that study motivated the use of the same formalism to study the molecular dynamics in MAPbBr$_3$.~\cite{Swainson2015,Brown2017}  Given the similarities between MAPbCl$_3$ and MAPbBr$_3$, we have opted to pursue the same approach here by dividing the EISF into two components, one representing the fraction of molecules that are constrained, therefore contributing no $Q$ dependence, and a second representing the fraction of molecules that are dynamic.  For the dynamic component, we have again assumed the simplest case where the motions are spherically symmetric.  In this case, the EISF takes the form

\begin{equation}
\begin{split}
I_{EISF}^{total}(Q)= \alpha {\tilde{A} \sin (Q r) \over {Qr}} + (1-\alpha),
\end{split}
\label{eqn:eisf}
\end{equation}

\noindent where $\tilde{A}$ is a factor that takes into account the effects of multiple scattering, which reduces the cross section in the limit $Q\rightarrow 0$.  This parameter varied from 0.9 to 1 for all temperatures.  It is thus indicative of the overall uncertainty associated with the EISF and was used to set the error bars in Figs.~\ref{fig:evol_temp} $(c-e)$.  The parameter $r$ is the radius associated with the molecular fluctuations; it was fixed to the ammonia hydrogen-hydrogen distance of 1.8\,\AA\ because a universal fit to all temperatures and momentum transfers found this value gives the best description of the data.  Representative fits are shown in Fig.~\ref{fig:evol_temp}, where panel $(a)$ illustrates how the quasielastic cross section varies with $Q$.  Panels $(c)$-$(e)$ show representative fits to Eqn.~\ref{eqn:eisf} for a series of three temperatures that indicate a good fit to the data.  At $T=30$\,K (Fig.~\ref{fig:evol_temp} $(c)$), the static or bound component of the quasielastic cross section dominates, and the scattering is independent of momentum transfer.  At higher temperatures (Fig.~\ref{fig:evol_temp} $(d-e)$) the bound component of the cross section accounts for less of the total scattering, and a decay with momentum is observed.

Fig.~\ref{fig:evol_temp} $(b)$ shows a plot of the dynamic fraction of the EISF as a function of $1000/T$.  In order to extract an activation energy for the molecules that are free to vibrate, these data were fit to an Arrhenius curve represented by the solid line and given by $\alpha=\alpha_0\exp{\left[ E_{\alpha}/k_BT \right]}$, where $\alpha_0=1.0 \pm 0.15$ and $E_{\alpha}=95 \pm 5$\,K.  These values may be compared to those determined for MAPbBr$_3$~\cite{Swainson2015}, where a similar analysis yielded $E_{\alpha}(Br)=51 \pm 10$\,K.  Physically, the parameter $\alpha_0$ should be equal to one because all molecules should become activated and contribute to the dynamics in the high temperature limit.  We found that fixing $\alpha_0$ for both the chlorine and bromine variants does not change the parameter $E_{\alpha}$ within our uncertainties.  Consequently, we conclude that the activation energy $E_{\alpha}$ decreases with increasing halogen atomic number.  This is consistent with the fact that the interstitial spaces in the heavier halide perovskites are larger, and thus one should expect the energy barrier to molecular activation to be correspondingly lower.  This was quantified in Ref. \onlinecite{Kieslich2014} by estimating an effect radii for the molecule with the goal of being able to calculate tolerance factors finding an estimate radii of 1.81, 1.96, and 2.20 \AA ~\cite{Saparov2016} for the chlorine, bromine, and iodine compounds, respectively.  The radii for the chlorine compound agrees well with the value we have obtained from our analysis outlined in Eqn. \ref{eqn:eisf} and also with previous analysis of the bromine compound.~\cite{Brown2017}

\begin{figure}
 \includegraphics[scale=0.45]{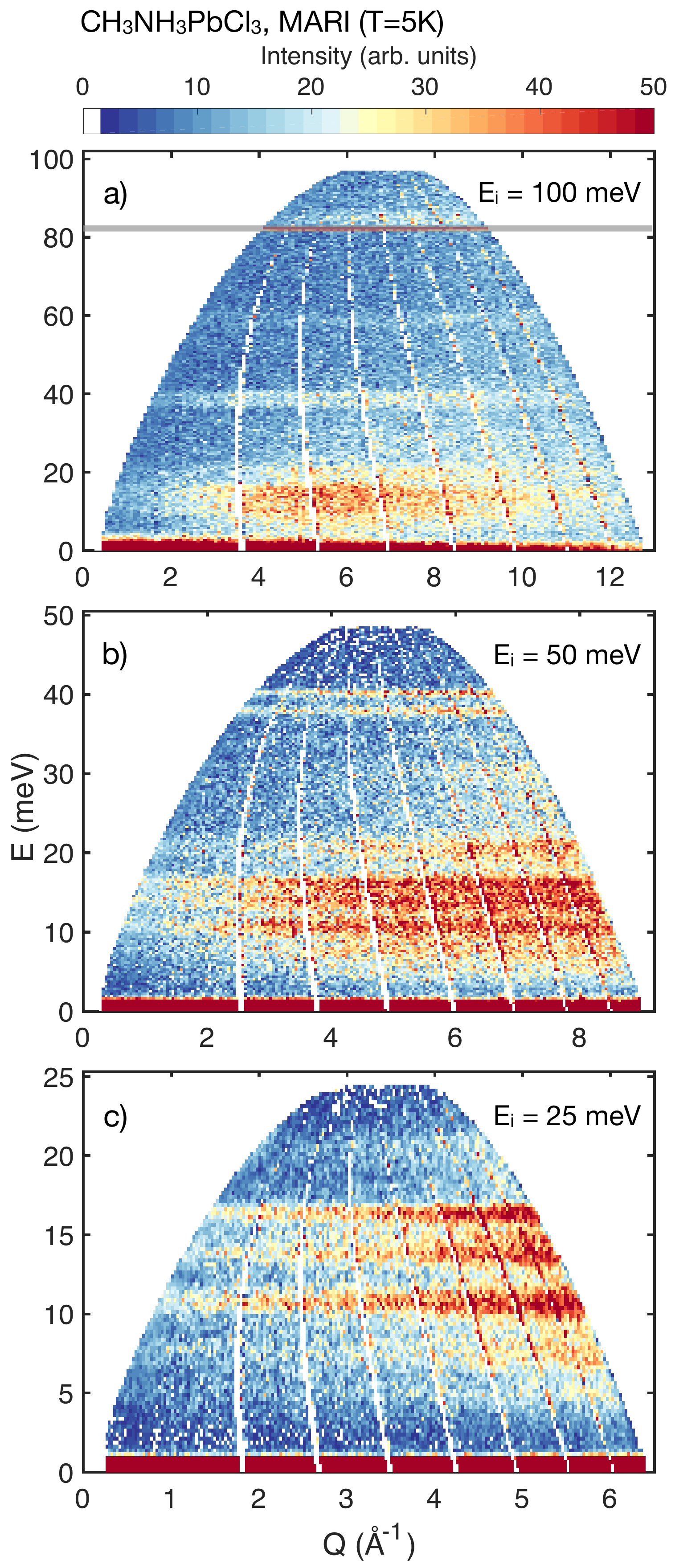}
 \caption{ \label{fig:mari} $(a)-(c)$ Phonon spectra measured by neutron inelastic scattering on MARI at $T=5$\,K for incident energies $E_i = 100, 50$ and 25\,meV, respectively. The data have been corrected for a background corresponding to the high temperature data. The grey line in $(a)$ corresponds to a spurion generated by the back side of the Gd Fermi chopper spinning at 600\,Hz.  This spurious feature is often referred to as the ``$\pi$-pulse."}
\end{figure}

\subsection{Molecular dynamics spectra at high energy}

The harmonic modes associated with the molecular cation are located at energies significantly higher than can be accessed by the dynamic range of $\pm 0.5$\,meV provided by the backscattering spectrometer IRIS.  To characterize these modes, neutron inelastic spectra spanning a much larger energy range were measured on the chopper instrument MARI using neutron incident energies of 100\,meV, 50\,meV, and 25\,meV, each of which covers a different dynamic range due to kinematic constraints.  Figure~\ref{fig:mari} $(a-c)$ displays neutron spectra for $E_i$ = 100, 50, and 25\,meV, respectively, measured at $T=5$\,K. These low temperature data (Fig.~\ref{fig:mari}) show five modes below 20\,meV, two modes around 40\,meV, one mode at 60\,meV, and another mode around 90\,meV, all of which agree with those identified by previous Raman and infrared studies of MAPbCl$_3$~\cite{Leguy2016bis, Glaser2015}.  Using the assignments of the Raman data reported in Ref.~\onlinecite{Leguy2016bis}, which were also compared to DFT calculations, we are able to identify the peaks measured on MARI with specific octahedral and molecular vibrations.

The phonon modes below 10\,meV correspond to transverse acoustic and optic phonons that are primarily associated with displacements of the PbCl$_6$ octahedra involving both twists (or rigid body motions) and distortions.  The peak near 11\,meV, termed the ``Nodding donkey" mode, corresponds to a rotational vibration of the cation around either the carbon or nitrogen atoms. These molecular motions are presumably affected by the octahedra tilting and distortions because of the hydrogen bonding with the halogen, and thus they provide some measure of the coupling between the molecule and the inorganic framework.  This is discussed in Refs.~\onlinecite{Swainson2015,Brown2017} for the bromine compound and in Refs.~\onlinecite{Quarti2014,Brivio2015} for MAPbI$_3$.  This coupling was also implicated in single crystal neutron scattering studies of the chlorine~\cite{Songvilay2018} and iodine~\cite{Parker2018} compounds where it was suggested that the molecular motions also affect the low-energy transverse acoustic phonons.

Above 12\,meV, the observed phonon modes correspond to lurching motions of the MA cations, and the mode at 60\,meV has been previously assigned to the torsions of the carbon-nitrogen (C-N) axis. Finally, a weak mode can be observed around 90\,meV that may arise from C-N bond stretching. These high-energy modes are associated with internal modes of the MA molecule and are consistent with previous Raman and infrared studies of the iodine and bromine compounds \cite{Leguy2016bis,Quarti2014}.

\begin{figure}
 \includegraphics[scale=0.38]{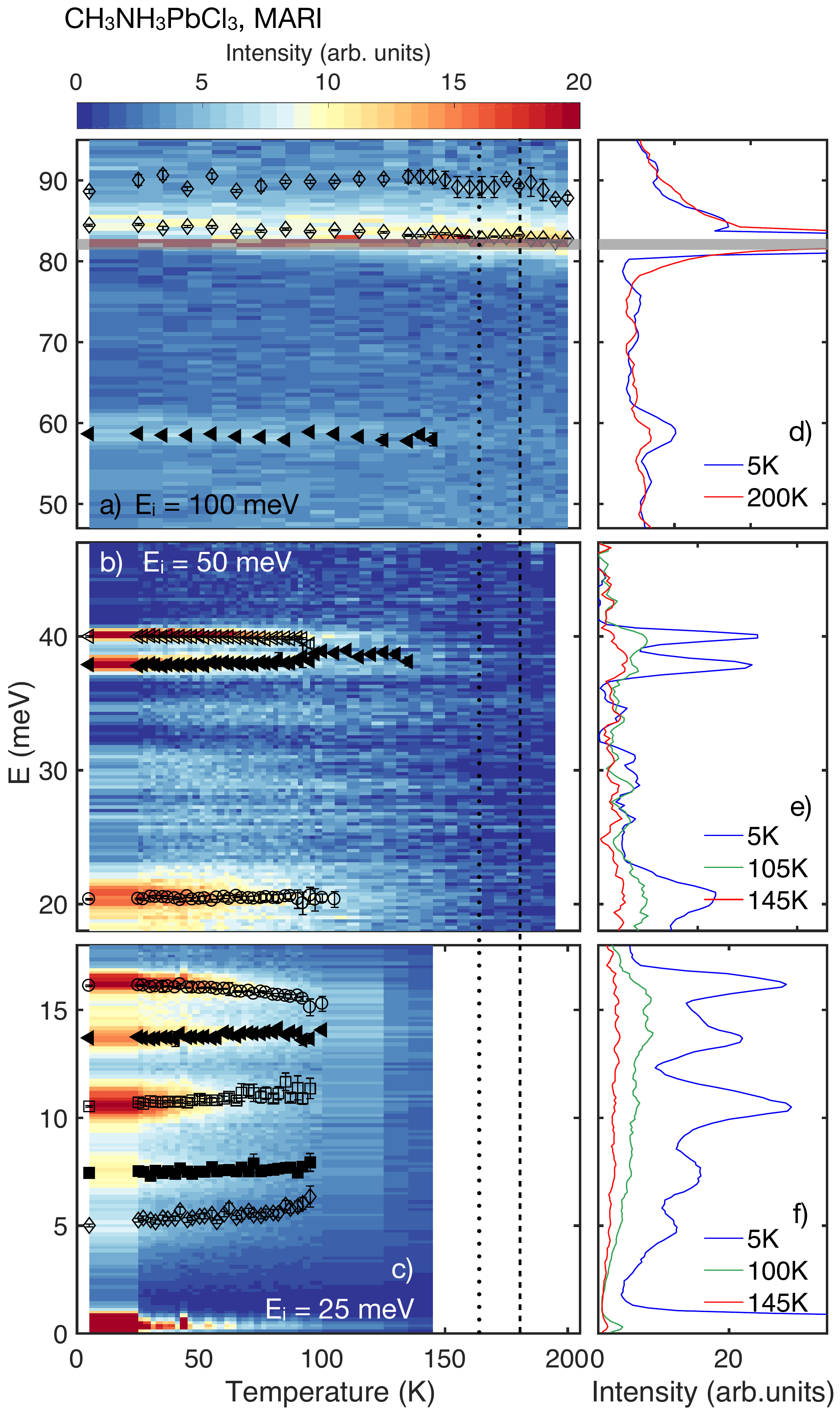}
 \caption{ \label{fig:mari_temp} $(a)-(c)$ Temperature dependence of the molecular modes in MAPbCl$_3$ measured at incident energies $E_i$ = 100, 50, and 25\,meV, respectively. At each temperature, the data was integrated over the entire $Q$ range, corrected for a background derived from the featureless high temperature data and corrected for the thermal factor. $(e)-(f)$ Energy cuts at different temperatures extracted from the $E_i$ = 100, 50, and 25\,meV data, respectively. The grey line in $(a)$ corresponds to a ``$\pi$-pulse" spurion from the back side of the Gd Fermi chopper spinning at 600\,Hz. }
\end{figure}

Figure~\ref{fig:mari_temp} $(a-c)$ shows color maps of the neutron inelastic scattering cross section that summarize the temperature dependence from 5\,K to 200\,K.  Fig.~\ref{fig:mari_temp} $(d-f)$ displays energy cuts of the $Q$-integrated spectra at different temperatures. These energy cuts reveal the surprising finding that, while the inelastic spectra exhibit sharp modes at 5\,K, the molecular vibrations become strongly overdamped (shorter lifetime) on heating $\sim$ 100\,K, which is far below either of the structural transitions associated with the inorganic framework (indicated by the vertical dashed lines).  This temperature dependence differs from that observed in the bromine compound, where a broad relaxational behavior is observed at temperatures that coincide with the structural transitions~\cite{Swainson2015} (see also Ref.~\onlinecite{Quarti2014,Brivio2015} for a discussion of MAPbI$_3$).  Internal motions, such as C-N distortions at high energies (Fig.~\ref{fig:structure}) persist above this temperature, however we note that these are internal molecular modes and hence not expected to couple to the inorganic framework surrounding the molecule.  These spectroscopic results therefore show that the molecular motions coupled to the framework become strongly damped at temperatures much lower than the structural transitions where the overall lattice distorts.

We finally note that the energy cuts in Fig.~\ref{fig:mari_temp} $(d-f)$ show that several modes around 15 and 40\,meV collapse into a single broad mode at 100\,K. This may be due to a change in the interactions between the hydrogen and chlorine atoms when the molecules start to reorient.  These results imply that the hydrogen bond tethering the molecules to the inorganic framework lattice is not a dominant energy scale and contrasts to the case of the bromine variant where the dampening of molecular vibrations occurs at a much higher temperature that coincides with the structural distortion of the lattice.  


\section{Discussion and Conclusion}


We observe a strong damping (reduction in lifetime) and weakening of the molecular modes in MAPbCl$_3$ on heating to $\sim$ 100\,K, a temperature that is much lower than that associated with the structural transitions of the inorganic framework near $\sim$ 170\,K.  By contrast, the molecular modes in the bromine compound become overdamped, and can be characterized by a relaxational lineshape, at temperatures that coincide with that where the overall structure distorts.~\cite{Swainson2015}  We observe that these two effects occur on very different temperature scales in MAPbCl$_3$, which suggests that the molecular and inorganic framework dynamics are actually relatively decoupled.  It is tempting to explain this result in terms of a halide-dependent hydrogen bond that is weaker for chlorine than for bromine.  However, recent first-principles calculations that show that the hydrogen bond energies in MAPbX$_3$ are relatively weak ($\sim$ 0.09\,eV/cation) and independent of halide for X = Cl, Br, and I.~\cite{Svane2017}  Our result is even more notable given the fact that the interstices of the framework cage surrounding the molecule in MAPbCl$_3$ are significantly smaller than those in MAPbBr$_3$ and MAPbI$_3$.

The softer molecular dynamics in MAPbCl$_3$, which are more sensitive to temperature than those in the other halide compounds, is reflected in our quasielastic measurements, which yield smaller activation energies from Arrhenius fits to the energy linewidths.  This is further consistent with calculations reported in Ref.~\onlinecite{Leguy2016bis}, which find a general softening of the lattice dynamics on going from iodine to chlorine organic-inorganic perovskites.

It is interesting to discuss this decoupling in terms of the electronic and thermal properties.  In the context of the importance of phonons, it has been proposed that high photovoltaic efficiency in the organic-inorganic perovskites is the result of ``Hot carriers"~\cite{Konig2010} where photo-excited charges can spatially propagate allowing time for them to be collected.  Normally, these charge carriers are highly damped by low-energy phonons, thereby reducing charge mobility.  However, in the organic-inorganic perovskites these phonons have such short lifetimes that they cannot affect the charge carriers, thus allowing highly efficient collection of the charge.

MAPbCl$_3$ generally shows a significantly lower mobility compared to MAPbl$_3$ for a broad range of sample geometries likely due to the larger band gap.~\cite{Herz2017} On adding chlorine to MAPbl$_3$, the diffusion lengths increase~\cite{Kim19:3,Stranks2013} through improving film quality.  It is difficult to understand the role of the molecular and framework lattice dynamics in improving the electronic properties.  The lead halide perovskites MAPbl$_3$~\cite{Parker2018}, MAPbBr$_3$~\cite{Ferreira2018}, MAPbCl$_3$~\cite{Songvilay2018}, and even fully inorganic CsPb(Br,Cl)$_3$~\cite{Fujii1974,Songvilay_unpub} all display quantitatively similar acoustic phonon lifetimes as a function of wave vector and temperature that results in poor thermal conductivity.~\cite{Kovalsky2017}  This alone illustrates that the low-energy acoustic phonon dampening is not the origin of the change in photovoltaic efficiencies with the change in halogen.  Another model is that the molecule increases the dielectric constant, thereby reducing the exciton binding energies.  However, electro-optics has reported very small binding energies of $\sim$ 2 meV~\cite{Lin2014}, which is well below the thermal energy of k$_{B}$T(=300 K) $\sim$ 25 meV making this mechanism unlikely.

We propose an alternative idea where the molecule does not participate in damping the low-energy acoustic phonons but instead provides a means of coupling the high energy charge sector to the much lower energy phonon sector.  A coupling between the $A$ site and the framework has been established through a comparison of CsPbBr$_{3}$ and MAPBr$_{3}$~\cite{Yaffe2017} and calculations have shown that such coupling is required to stabilize an orthorhombic unit cell at low temperatures~\cite{Bechtel2018} as is present in the fully inorganic variants.  While these two channels exist on very different energy scales, calculations have shown that molecular motions coupled with the underlying inorganic framework do influence the electronic structure~\cite{Motta2015}.  We also note that internal molecular modes occur on much higher energy scales than do overall lattice fluctuations (like those observed in the perovskite SrTiO$_3$~\cite{Cowley1964}) and may provide a means of coupling the electronic sector in organic-inorganic perovskites to the already highly-damped low-energy acoustic phonon response.  This maybe consistent with the enhanced dissociation of charge transfer states with molecular loading of the framework lattice.~\cite{Bernardo2014,Leblebici2017}  In this heuristic picture, the strength of this coupling between electronic degrees of freedom and the low-energy lattice dynamics would depend on the band gap which is set by the lead-halide bond and is expected (based on the unit cell volumes) to be smaller for the iodine and bromine variants.~\cite{Walsh2015,Bardeen1950}   This idea would imply that fully inorganic perovskites have the potential to be efficient photovoltaics, provided that a means of coupling the phonons on the meV scale with electronic excitations on the eV scale can be achieved.   The speculation of framework-molecule coupling, set by the lead-halide bond and hence the band gap, being the important feature for attractive electronic properties requires further theoretical and experimental investigation.

We now discuss the origin of the dampening of the molecular motions in the chlorine compound compared to the bromine variant.  We have shown that damped, or lifetime shortened, molecular dynamics occur at a significantly lower temperature than in the bromine compound published previously.  We have used this fact combined with the common acoustic phonon dampening observed in both organic-inorganic and fully inorganic perovskites to question the role of the molecule in enhancing the electronic properties.  As discussed above, the volume that is available for the molecule to rotate can be quantified by an effective radius and is noted to be smaller in the chlorine variant than the bromine and iodine organic-inorganic perovskites.   This decreased volume around the molecule may make it more sensitive to disorder that is present in the surrounding inorganic cage.  The highly damped acoustic phonons for shorter wavelengths located away from the Brillouin zone center maybe indicative of disorder in the low temperature inorganic framework structure.  This is present for all halogens, but the molecular motions in the chlorine variant maybe more susceptable to this disorder owing to the smaller volume for the molecular motions to occur.   This is speculation at this point based on the dynamics and future work comparing the structure through improved diffraction experiments will be required to corroborate this suggestion.  However, we note that recent work on fully inorganic CsPbBr$_{3}$ did note significant disorder through enhanced thermal factors of the atomic positions.~\cite{Songvilay_unpub}

In summary, we report a neutron inelastic scattering investigation of MAPbCl$_3$ that was motivated by recent studies on the photovoltaic properties of organic-inorganic perovskites and neutron studies performed on the bromine and iodine variants.  We observe a decoupling of the molecular and  framework dynamics based on the temperature dependence of the molecular vibrations which differ from the response of the distorted framework driven by acoustic phonons.  We therefore speculate that the molecule may provide a means of connecting the disparate energy scales of the charge and low-energy phonon sectors, which in turn may contribute to the attractive electronic properties of these materials.    In this picture, the important energy scale is set by the lead-halide bond, and not by hydrogen bonding. 

\begin{acknowledgments}
We acknowledge funding from the EPSRC and the STFC. We are thankful to M. A. Green, I. Swainson, and P. Bourges for fruitful discussions.  This work was also supported by the Natural Sciences \& Engineering Research Council of Canada (NSERC Grant No. 203773) and the U. S. Office of Naval Research (ONR Grant No. N00014-16-1-3106).
\end{acknowledgments}

\end{document}